\documentclass[letterpaper, 10 pt, conference]{ieeeconf}  % Comment this line out
\IEEEoverridecommandlockouts                              % This command is only
\usepackage[utf8]{inputenc}
\usepackage[T1]{fontenc}
\usepackage{graphicx}
\usepackage{amsmath}

\usepackage[ruled,vlined]{algorithm2e}

\usepackage{amsmath}
\usepackage{authblk}
\title{\LARGE \bf
Posterior Cramer-Rao Lower Bound based Adaptive State Estimation for Option Price Forecasting
}

\author{Kumar Yashaswi$^{1}$% <-this % stops a space 
\affil{Department of Mathematics, Indian Institute of Technology Kharagpur}
\thanks{*This work was supported by Department of Mathematics, Indian Institute of Technology Kharagpur}% <-this % stops a space
\thanks{$^{1}$K. Yashaswi- Department of Mathematics, Indian Institute of Technology Kharagpur}%
}

\begin{document}

\maketitle
\thispagestyle{empty}
\pagestyle{empty}

%%%%%%%%%%%%%%%%%%%%%%%%%%%%%%%%%%%%%%%%%%%%%%%%%%%%%%%%%%%%%%%%%%%%%%%%%%%%%%%%
\begin{abstract}

The use of Bayesian filtering has been widely used in mathematical finance, primarily in Stochastic Volatility models. They help in estimating unobserved latent variables
from observed market data. This field saw huge developments in recent years, because of the increased computational power and increased research in the model parameter estimation and implied volatility theory. In this paper, we design a novel method to estimate underlying states (volatility and risk) from option prices using Bayesian filtering theory and Posterior Cramer-Rao Lower Bound (PCRLB), further using it for option price prediction. Several Bayesian filters like Extended Kalman Filter (EKF), Unscented Kalman Filter (UKF), Particle Filter (PF) are used for latent state estimation of Black-Scholes model under a GARCH model dynamics. We employ an Average and Best case switching strategy for adaptive state estimation of a non-linear, discrete-time state space model (SSM) like Black-Scholes, using PCRLB based performance measure to judge the best filter at each time step [1].
Since estimating closed-form solution of PCRLB is non-trivial, we employ a particle filter based approximation of PCRLB based on [2]. We test our proposed framework on option data from S$\&$P 500, estimating the underlying state from the real option price, and using it to estimate theoretical price of the option and forecasting future prices. Our proposed method performs much better than the individual applied filter used for estimating the underlying state and substantially improve forecasting capabilities.

\vspace{5mm} %5mm vertical space

Keywords- Option Theory; Stochastic volatility; Bayesian Filtering; Particle Filter; Posterior Cramer-Rao Lower Bound (PCRLB) 

\end{abstract}

%%%%%%%%%%%%%%%%%%%%%%%%%%%%%%%%%%%%%%%%%%%%%%%%%%%%%%%%%%%%%%%%%%%%%%%%%%%%%%%%
\section{INTRODUCTION}
Option pricing theory involves obtaining a theoretical price of an option contract on an asset using several factors like asset price, strike price, volatility, time to expiration, interest rate, etc. These estimated prices are widely used for improving investment yield and hedging purposes for portfolios. Black-Scholes equation derived in 1973 [3] laid the foundation for pricing option's with the following works improving upon it. With a lot of factors governing option prices, an effective option price forecasting tool is important for an investor to identify profitable trades. Stochastic volatility models are one such set of models that model the movement of option prices.
 GARCH model is one such class of volatility model that is used significantly in option pricing. Though it does not come under the stochastic volatility models due to computing deterministic value, we add a noise term to it to make it a stochastic quantity. GARCH model dynamics is used to model volatility along with Black-Scholes (BS) model for estimating option prices and make a forecasting framework of option prices. \par
BS model requires calibration of its parameters (volatility, risk-rate) based on observed market data like asset prices and option prices. A popular technique for calibration of parameters is non-linear filtering which has been widely used in finance for yield curve modelling, time-series filtering and stochastic volatility modelling [4], [5]. In Bayesian filtering theory, specific model dynamics is defined for a given system, and the underlying state or signal is estimated using concepts of recursive Bayesian estimation. Bayesian methods in the industry have been described thoroughly in [6]. This is used to determine unobserved latent variables from observed market data under given model dynamics like GARCH and BS. Volatility and risk which are the parameter's of BS model are taken as the state-space for filters. \par

Since BS model is a non-linear function of underlying asset prices, Kalman filter can't be used since it requires linear and Gaussian conditions of measurement and state space. Hence we use three non-linear filtering techniques EKF, PF, UKF for latent state estimation. One major drawback with using individual filters is that no single filter will be optimal for all time steps. At different time step, the observed option prices may be more accurately be modelled by different filters under given model dynamics. \par
In this paper, we use a framework for adaptive state estimation using filter switching strategy based on Posterior Cramer-Rao Lower Bound (PCRLB) measure. PCRLB gives a theoretical lower bound of mean square error (MSE) of a non-linear Bayesian filter.
[1] proposed an efficient strategy to select the most optimal filter at each time step from a set of Bayesian filters for a general non-linear, stochastic, discrete-time state-space models (SSMs). Their idea was based on the fact that there is no single Bayesian estimator that will perform optimally for a given non-linear system under all operating conditions. The switching strategy used for state estimation was to switch between a set of Bayesian filters at each time step based on their performance. The performance metric was derived using PCRLB, which is defined for a class of non-linear SSMs. \par
The computation of a closed-form solution of PCRLB itself is a very difficult task and is not possible unless a true state is known, which for the option pricing theory is missing in form of volatility and risk, a non-observable quantity. Hence we use a particle filter based approximation of PCRLB based on the work of [2]. The computation of PCRLB along with the use of adaptive state estimation strategy from [1] gives a stable framework for estimating the underlying states of BS model. [1] proved the stability of the adaptive switching strategy and [2] justified the accuracy of particle filter based approximation of PCRLB as it ensured that the proposed numerical solution does not result in accumulation of errors. \par
We largely modify work done by [7], [8], and [9]. The major addition to their work being the adaptive state estimation strategy employed which we shall compare with individual filters. We test the method on S$\&$P 500 options data which provide a large amount of highly liquid data for training and test dataset. We use a 1 time step forecasting technique on test data using the filter estimated in previous time step from the proposed strategy.

To the best of our knowledge, this is the first of many work that uses an adaptive state estimation technique for inferring underlying state from observed option prices movement, and further using it for option price forecasting. The main aim of our work is to investigate the efficiency of using the switching strategy to optimally and sequentially select a different Bayesian filter at each time step for stochastic volatility models and compare its performance with respect to individual filters throughout the training data. Such a framework can be used to design an effective option trading strategy.

\subsection{Related Work}
With the increased research being done on Bayesian filtering theory for several applications, their use in state estimation of stochastic volatility models has increased widely over the past years. [9], [10] were one of the major works in this domain. The usage of filtering theory was aided using their research. [8] used a hybrid model of EKF and SVM for option-price prediction, built on Black-Scholes formula. [7] further improved upon it using UKF. [4] and [5] gives a basic overview of filtering theory in finance including in stochastic volatility model. \par
Several works have been done for estimating PCRLB, which is a vital component computed for our proposed method. [11] used a similar approximation to [2] for PCRLB and tested it on non-linear tracking applications. Additionally [12] evaluated the bound for three cases: recursive estimation of slowly varying parameters of an autoregressive process, tracking a slowly varying frequency of a single cisoid in noise and tracking parameters of a sinusoidal frequency with sinusoidal phase modulation. An alternate Bayesian filter switching strategy was used by [13], using a performance-based filter switching strategy for state estimation under known model parameters case. 

\section{Background}

We discuss computation of PCRLB which is the main concept on which the switching strategy is based. For theory on Filtering techniques used, readers are referred to [14], [15], [16].

\subsection{Posterior Cramér-Rao lower bound}
Cramer–Rao bound (CRB) is a commonly used statistical lower bound in time-invariant models. It is used to indicate the lower theoretical bound of the mean square error (MSE) of a state-estimator.
In time-varying systems such as the option pricing framework, the estimated price movement is considered random since it corresponds to an underlying nonlinear, randomly driven state process. For such systems a Posterior Cramer–Rao bound (PCRLB) is computed. Since option pricing is defined in non-linear terms, we are concerned with using PCRLB for lower-bound analysis of the MSE of our SSM and helps in the algorithm for adaptive state estimation of options price movement according to BS and GARCH model formulation. Hence it's an important statistic to compute for the state estimation system. Two of the most vital uses of PCRLB is 
\begin{itemize}
\item gives a benchmark measure of fit for different non-linear filters
\item performance comparison of several non-linear filters against one another and to that of an optimal filter
\end{itemize}
For a system of non-linear model:
\begin{equation}
X_{t+1} = f_{t}(X_t)+V_t,\\
\hspace{0.1cm} V_t \sim N(0,Q_t) 
\end{equation}
\begin{equation}
Y_{t} = g_{t}(X_t)+W_t,\\
\hspace{0.1cm} W_t \sim N(0,R_t)
\end{equation}
the PCRLB inequality is described as:
\begin{equation}
P_{t|t} \equiv E_{p(X_{0:t},Y_{1:t})}[(X_t - {\widehat{X}}_{t|t})(X_t - {\widehat{X}}_{t|t})^T] \geq J_t^{-1}
\end{equation}
For a randomly sampled measurement sequence $\{Y_{1:t}\}_{t \epsilon N}$ from Eq (1-2), the MSE of any Bayesian filter at ${t \in N}$ is bounded where $P_{t|t}$ is a $s\times s$ sized MSE matrix which belongs to the set of symmetric positive definite matrices $S_{++}^s$, $X_{t|t}\equiv X_{t}(Y_{1:t}): R^{tm}\rightarrow R^s$ is point estimate of state $X_t$ at time $t$. The $s\times s$ matrix $J_t$ is known as the posterior Fisher information matrix (PFIM) is a way of measuring the information present in a state $X$ using a distribution that models $X$. It's inverse is the PCRLB matrix. The proof of the above equation can be seen in works like [17]. 
Eq (3) can also be written in the form 
\begin{equation}
P_{t|t}^S \equiv E_{p(X_{0:t},Y_{1:t})}[\|X_t - {\widehat{X}}_{t|t}\|^2] \geq Tr[J_t^{-1}]
\end{equation}
where $Tr$ is the trace of matrix and $\|.\|$ is the 2-norm. Eq (42) can be attributed to the fact that $P_{t|t}-J_t^{-1}\geq0$ is a positive definite matrix for all state estimates at time $t\in N$.
The model in Eq (1-2), following set of assumptions (Assumption (3.2-3.5) as described in [1]), gives way to the recursive computation of the PFIM as: 
\begin{equation}
J_{t+1} = D_t^{22} - [D_t^{12}]^T (J_{t}+D_t^{11})^{-1}D_t^{12}
\end{equation}
The individual components are defined as in Eq (45-47) below:
\begin{equation}
D_t^{11} = E_{p(X_{0:t+1},Y_{1:t+1})}[-\triangle_{X_{t}}^{X_{t}} log(p(X_{t+1}|X_t))]
\end{equation}
\begin{equation}
D_t^{12} = E_{p(X_{0:t+1},Y_{1:t+1})}[-\triangle_{X_{t}}^{X_{t+1}} log(p(X_{t+1}|X_t))]
\end{equation}
\begin{equation}
\begin{split} 
D_t^{22} = E_{p(X_{0:t+1},Y_{1:t+1})}[-\triangle_{X_{t+1}}^{X_{t+1}} log(p(X_{t+1}|X_t)) \\- \triangle_{X_{t+1}}^{X_{t+1}} log(p(Y_{t+1}|X_{t+1}))]
\end{split}
\end{equation}

The gradients using the Laplacian operator $\triangle$ is calculated at the true
states. The PFIM matrix is defined for initial states as:
\begin{equation}
J_{0} = E_{p(X_{0})}[-\triangle_{X_{0}}^{X_{0}} log(p(X_{0})]
\end{equation}

Any Bayesian estimator of SSM model that solves for $p(X_{t}|Y_{1:t})$ is valid to obtain a PCRLB inequality.  
Getting a closed-form solution to the PFIM or PCRLB is non-trivial, due to the complex integrals involved in Eq (3). To solve this issue, [2] used particle filter to give approximate solution to PFIM, hence PCRLB solving the set of Eq (6-8) using particle filter. We add a modification to the approach used in [2] by adding the information present in state estimated through different filters.

\subsection{Particle Filter Approximation of PCRLB}
Since Black-Scholes is a  non-linear SSM with additive Gaussian noise, we work with the model represented by Model 4.15 in [2], which gives an appropriate approximation for the set of systems. This technique was inspired as an alternative to the Monte-Carlo methods which required an ensemble of true state's and measurements which are not available in many defined systems.
We shall explain the general approach and result without jumping to the derivation, which is rigorously explained in [2]. 
Based on the approach mentioned in [2], approximation of Eq (6-8) was derived as:
\begin{equation}
D_t^{11} = \frac{1}{MN}\sum_{j=1}^{M}\sum_{i=1}^{N}[\nabla_{X_{t}} f_{t}^T(X_{t|t+1}^{i,j})]Q_{t}^{-1}[\nabla_{X_{t}} f_{t}(X_{t|t+1}^{i,j})]
\end{equation}
\begin{equation}
D_t^{12} = \frac{1}{MN}\sum_{j=1}^{M}\sum_{i=1}^{N}-[\nabla_{X_{t}} f_{t}^T(X_{t|t+1}^{i,j})]Q_{t}^{-1}
\end{equation}
\begin{equation}
\begin{split} 
D_t^{22} = Q_{t}^{-1} + \frac{1}{MN}\sum_{j=1}^{M}\sum_{i=1}^{N}([\nabla_{X_{t+1}} g_{t+1}^T(X_{t+1|t}^{i,j})]\\ R_{t+1}^{-1}[\nabla_{X_{t+1}} g_{t+1}(X_{t+1|t}^{i,j})]) 
\end{split}
\end{equation}
The particles are N-samples following $\{X_{t+1|t}^{i,j}\}_{i=1}^{N} \sim p(x_{t+1}|y_{1:t}^j)$ and $\{X_{t|t+1}^{i,j}\}_{i=1}^{N} \sim p(x_{t}|y_{1:t+1}^j)$, with M measurement sequence being $\{Y_{1:t+1}=y_{1:t+1}\}_{i=1}^{M}$ obtained from historical data. For the problem of option pricing there is only 1 measurement sequence. \par
The distribution $p(x_{t+1}|y_{1:t}^j)$ and $p(x_{t}|y_{1:t+1}^j)$ are obtained through distribution $p(dx_{t:t+1}|y_{1:t+1})$, which can be computed through $p(dx_t|y_{1:t})$ and $p(dx_{t+1}|y_{1:t+1})$ as explained in [2]. Firstly, $p(dx_t|y_{1:t})$ and $p(dx_{t+1}|y_{1:t+1})$ are obtained by Algorithm (1).
\begin{algorithm}
\SetAlgoLined
\KwResult{Particles from approximated probability distribution's $p(dx_t|y_{1:t})$ and $p(dx_{t+1}|y_{1:t+1})$}
 \textbf{Initialization}: Estimated State $x_{t|t}$ and state Covariance $P_{t|t}$ through filter $f \in B$ where $B$ is set of Bayesian Filters; \\
 \textbf{if} {t<T}{ \;
  \textbf{Sampling Step}-\;
  1) For i=1,...N sample $(X^{(i)}_{t|t}) \sim p(x_{t|t},P_{t|t})$\;
  2) For i=1,...N compute $(X^{(i)}_{t+1|t}) \sim \frac{1}{N}\sum_{i=1}^{N}p(X_{t+1|t}| X^{(i)}_{t|t})$- according to state equation (56)\;
  3) For i=1,...N set particle weight's $w^{(i)}_{t+1|t}= \frac{1}{N}$ \;
  4) For i=1,...N, $\{Y_{t+1} = y_{t+1}\}$ set particle weight's $w^{(i)}_{t+1|t+1}= w^{(i)}_{t+1|t} \frac{p(Y_{t+1} | X^{(i)}_{t+1|t})}{\sum_{j=1}^{N}p(Y_{t+1} | X^{(j)}_{t+1|t})}$ \;
  5) Resample particle set $\{X^{(j)}_{t+1|t+1}\}$ for j=1,..N with replacement from $\{X^{(i)}_{t+1|t}\}$ using resampling technique such that \
  $Pr(X^{(j)}_{t+1|t+1} = X^{(i)}_{t+1|t}) = w^{(i)}_{t+1|t+1}$, where Pr is probability measure. Set \{$W^{i}_{t+1|t+1} = \frac{1}{N}$\} \;
  \textbf{Output: Particle Filter Approxmated probability distribution}- $ X^{(i)}_{t|t} \sim p(dx_t|y_{1:t})$ and $ X^{(i)}_{t+1|t+1} \sim p(dx_{t+1}|y_{1:t+1})$ for i=1,2,..N\;
 }
 \caption{Particle Filter Algorithm for generating PCRLB distributions}
\end{algorithm}

Most of the algorithm is similar to the one used in [2]. The major variation was the use of state and covariance matrix obtained from the set of individual Bayesian filters as opposed to only particle filter. Based on the output, the SMC approximation of target distribution $p(dx_{t:t+1} | y_{1:t+1})$ can be defined as:
\begin{equation}
\widetilde{p}(dx_{t:t+1} | y_{1:t+1}) = \sum_{i=1}^{N} W^{i}_{t|t,t+1|t+1}\delta_{X^{i}_{t|t},X^{i}_{t+1|t+1}}(dx_{t:t+1})
\end{equation}
where
\begin{equation}
W^{i}_{t|t,t+1|t+1} = \frac{\zeta^{i}_{t|t,t+1|t+1}}{N\sum_{j=1}^{N}\zeta^{j}_{t|t,t+1|t+1}}
\end{equation}
\begin{equation}
\zeta^{i}_{t|t,t+1|t+1} = \frac{p(X^{i}_{t+1|t+1}|X^{i}_{t|t})}{N\sum_{m=1}^{N}p(X^{i}_{t+1|t+1}|X^{m}_{t|t})}
\end{equation}
and $\delta_{X^{i}_{t|t},X^{i}_{t+1|t+1}}(.)$ is marginalized Dirac delta function in $dx_{t:t+1}$, centred around the random particle's $X^{i}_{t|t},X^{i}_{t+1|t+1}$. The particles are resampled according to:
\begin{equation}
Pr(X^{(j)}_{t:t+1|t+1} = \{X^{(i)}_{t|t};X^{(i)}_{t+1|t+1}\}) = W^{(i)}_{t|t,t+1|t+1}
\end{equation}
where $\{X^{i}_{t:t+1|t+1}\}^{N}_{i=1}$ are resampled i.i.d particles. After resampling target distribution is approximated as:
\begin{equation}
\widetilde{p}(dx_{t:t+1} | y_{1:t+1}) = \frac{1}{N}\sum_{i=1}^{N} \delta_{X^{i}_{t:t+1|t+1}}(dx_{t:t+1})
\end{equation}

According to Lemma (4.13) in [2], the SMC based approximation of $p(dx_t | y_{1:t+1})$ is given by :
\begin{equation}
\widetilde{p}(dx_{t} | y_{1:t+1}) = \frac{1}{N}\sum_{i=1}^{N} \delta_{X^{i}_{t|t+1}}(dx_{t})
\end{equation}
$\delta_{X^{i}_{t|t+1}}(.)$ is marginalized Dirac delta function in $dx_{t}$, centred around the random particle $X^{i}_{t|t+1}$. Distribution $\widetilde{p}(x_{t+1}| y_{1:t})$ is approximated similarly as:
\begin{equation}
\begin{split}
\widetilde{p}(dx_{t+1}| y_{1:t}) =\int_{\chi} p(x_{t+1}|x_{t})\widetilde{p}(dx_{t}|y_{1:t}) = \\ \frac{1}{N}\int_{\chi} p(x_{t+1}|x_{t})\sum_{i=1}^{N} \delta_{X^{i}_{t|t}}(dx_{t}) \\ 
= \frac{1}{N}\sum_{i=1}^{N} p(X_{t+1|t+1}|X^{m}_{t|t})
\end{split}
\end{equation}

The particle filter approximation of both PFIM and PCRLB using Eq (10-12) and matrix inversion lemma are given by:
\begin{equation}
J_{t} = D_t^{22} - [D_t^{12}]^T (J_{t}+D_t^{11})^{-1}D_t^{12}
\end{equation}
\begin{equation}
\begin{split}
J_{t}^{-1} = [D_t^{22}]^{-1} - [D_t^{22}]^{-1}[D_t^{12}]^T \times [D_t^{12}[D_t^{22}]^{-1} [D_t^{12}]^{T} \\ - (J_{t}+D_t^{11})]^{-1} D_t^{12}[D_t^{22}]^{-1}
\end{split}
\end{equation}
where $J_{t}$ and $J_{t}^{-1}$ are particle filter approximated PFIM and PCRLB respectively. The convergence of particle filter approximation to true PCRLB can be shown. It is not necessarily implied about the convergence of the SMC based PCRLB and MSE to its theoretical values, it does provide a good theoretical basis for the numerous approximations.

\subsection{Adaptive State-Estimation Switching Strategy}
We have discussed the relevant concepts that would be required for designing a strategy to estimate state-space from stochastic volatility models. It will select different Bayesian estimators to be used in each step, depending on the optimality of each filter. PCRLB judges the MSE performance of a Bayesian estimator, with the estimator with the closest MSE to PCRLB is the best performing. MSE matrix is computed for each filter and PCRLB through particle filter approximation explained in the last subsection. The strategy is initiated with a set of Bayesian estimators, and switch between them as and when
required based on their performance. To get a scalable metric we define performance metric as:
\begin{equation}
\Phi_t = J_t^{-1}oP_{t|t}^{-1}
\end{equation}
$\Phi_t \in R^{s\times s}$ is the new performance matrix to assess the performance of the Bayesian state estimator used, where $s$ is the number of states to be estimated.  An alternative definition of equation (22) is:
\[
\Phi_t (i,j) =
\begin{cases}
J_t(j,j)^{-1}[P_{t|t}(j,j)]^{-1}, &\text{if i=j;}\\
0, &\text{if i $\neq$ j.}\\
\end{cases}
\]
where $i,j \in 1,2,..s$. We state some results from [1] with given proof in the same-
\begin{equation}
0< \Phi_t(j,j) \leq 1
\end{equation}

This lead's to the condition $Trace[\Phi_t] \in (0,s]$ for all Bayesian estimators at any time. This puts a bound on the performance measure as compared to PCRLB which may be unbounded in many applications.
For the strategy, the best point estimate at each time step is decided based on the maximum value of performance metric. The adaptive state estimation strategy is based on using (22) to judge the best state estimator $f\in B$ at time step $t$. Our initial process is computing performance metrics using MMSE and PCRLB matrix for all Bayesian estimators from $B$. We then switch between them for each time step based on which gave the best performance. We employ the Average and Best case switching strategy from [1] for our state estimation. The average-case strategy can be described as:
\begin{itemize}
\item Input: Set of measurements upto time period t, set of Bayesian filters $B$
\item Step 1: Compute MMSE matrix, PCRLB and Performance metric for all Bayesian estimator $f\in B$
\item Step 2: Compute $Trace[\Phi_t^f]$ using Performance metric for all $f\in B$
\item Step 3: Solve for f such that $\widehat{f} = arg max_{f}(Trace[\Phi_t^f])$ 
\item Output: $\widehat{X}_{t|t}^{(\widehat{f})}$ is the point estimate to be selected at time $t$ 
\end{itemize}
and the best-case switching strategy as:
\begin{itemize}
\item Input: Set of measurements upto time period t, set of Bayesian filters $B$
\item Step1: Compute MMSE matrix, PCRLB and Performance metric for all Bayesian estimator $i\in B$
\item Step 2: Compute $Trace[\Phi_t^f]$ using Performance metric for all $f\in B$
\item Step 3: Set j <- 1, While j $\leq$ s, repeat, \\
1) Solve for f such that \\
$\widehat{f} = arg max_{f}(Trace[\Phi_t^f](j,j))$ \\
2) Select $\widehat{X}_{t|t}^{(\widehat{f})}(j)$ as point estimate\\
3) j <- j+1 
\item Output: $\widehat{X}_{t|t}$ is the point estimate to be selected at time $t$ 
\end{itemize}
The major difference to average-case, which is a one-dimensional switching strategy, best-case strategy provides an s dimensional strategy, where each state has its switching strategy defined.\par  
All the estimators in the set estimate all the states of the system at each sampling time parallelly, but only the ones suggested by the switching strategy are selected as the final estimate. This technique gives the advantage of parallel computation for increasing computation of the sub-optimal state.
The numerical stability of the algorithm is explained thoroughly in [1] by applying the boundedness of stochastic processes.

\section{Problem Formulation}

\subsection{State Space Modelling of Stochastic Volatility Model}
We model the option prices using Black-Scholes model. 
The initial formulation of BS model assumed constant volatility structure. A non-linear behaviour was observed in volatility that came to be known as volatility smile, which has since been observed for all asset classes. \par
To get a structure of volatility, we combined GARCH(1,1) with Black-Scholes model and formulate it in the dynamics of the state-space model. 
The Black-Scholes price for call and put option prices ($C$ and $P$ respectively) are represented as:
\begin{equation}
C = SN(d_1)- Ke^{-r\tau}N(d_2)
\end{equation}
\begin{equation}
P = -SN(-d_1)+ Ke^{-r\tau}N(-d_2)
\end{equation}
where $S$ is the stock price, $K$ is strike price, $r$ is the risk-free rate, $\tau$ is time left to expiration. $d_1$ and $d_2$ are defined as:
\begin{equation}
d_1 = \frac{ln(S/K) + (r + \sigma^2/2)\tau}{\sigma\sqrt{\tau}}
\end{equation}
\begin{equation}
d_2 = d_1 - \sigma\sqrt{\tau}
\end{equation}

The GARCH model assumes the underlying stock price $S_t$ has a stochastic variance $v_t$, which is estimated using predicted variance at $t-1$ step, long-run average variance rate
$V_L$ and stock returns up to $t$ ($u_t$). The dynamics are represented as:
\begin{equation}
v_t = \omega + \alpha u_{t}^2 + \beta v_{t-1} + w_t
\end{equation}
\begin{equation}
r_t = r_{t-1} + v_t
\end{equation}
\begin{equation}
c_t = BS(v_t,r_t) + z_t
\end{equation}
where Eq (28) is the GARCH(1,1) formulation with added Gaussian noise $w_t$. We add noise variation's in risk-free rate as well, working with a combination of formulation in both [8] and [9]. Eq (28-29) represents the state equations, and Eq (30), which is the BS prices with a noise term $z_t$, represents the measurement equation. 
This will help us to employ the adaptive state estimation technique to probabilistically estimate the volatility state of the dynamic system from the noisy option price observations. \par 
Algorithm (2) explains the process of state estimation. The $filterUpdate$ is the prediction and update step of each individual filter. It is then followed by the PCRLB computation in $PCRLB$ for all $f \in B$. These computations are then used in the switching strategy function $SwitchingStrategy$ for selecting the most optimal state estimated from all filter. The additional function parameters are represented as $\theta$ and $\Theta$.

\begin{algorithm}
\SetAlgoLined
\KwResult{Estimated underlying states $\widehat{x}_{t}$ using option prices $c_t$ for $t \in \{1,2,3,..T\}$} 
 \textbf{Initialization}: Initial State $x_{0} = \{v_0,r_0\}$, state covariance $P_{0}$ and $J^{f}_{0} = P^{-1}_{0}$ for filters $f \in B$ where $B$ is set of Bayesian filters;\\
 \While{$t < T$}{
  \While{$f \in B$}{
$({x}^f_{t}, P^f_{t}) \leftarrow filterUpdate^f(c_t, {x}_{t-1}, P_{t-1}, \theta_t)$;
$(J^{f}_{t+1}, (J^{f}_{t+1})^{-1}) \leftarrow PCRLB(J^{f}_{t}, c_{t+1}, {x}^f_{t}, P^f_{t}, \Theta_{t+1})$;
  }
  $({x}_{t}, P_{t}) \leftarrow SwitchingStrategy(({J}^B_{t})^{-1},{x}^B_{t}, P^B_{t})$
  }
  \textbf{Output: Volatility and Risk state estimated through the switching strategy}- 
  $ X^{(i)}_{t|t} \sim p(dx_t|y_{1:t})$ and $ X^{(i)}_{t+1|t+1} \sim p(dx_{t+1}|y_{1:t+1})$ for i=1,2,..N
 \caption{Adaptive State Estimation Strategy}
\end{algorithm}

\subsection{Option Data Set}

In the previous sections, we had described our proposed technique based on PCRLB measure and the filtering techniques for estimation of underlying volatility and risk state under a Black-Scholes and GARCH(1,1) modelling framework. We will use 3 different filters (EKF, UKF, PF) for state estimation and particle filter for approximate computation of PCRLB. 

The data used for backtesting the proposed method is the S$\&$P 500 index option [18] from the year 2019 to 2020. For each day, it has option quotes for both call and put options with various expiration and strike price data. 

For backtesting forecasting accuracy we consider roughly 150 daily call option prices. We took a single expiration date of December 2020 based on the liquidity of options with strike prices of 2000, 2500 and 3000. 
The predictions are made for 1 day ahead and as the new option prices come in, the predicted values are corrected and PCRLB is computed according to the new prices. This ensures the error in forward prediction by the filters do not add up. \par
We further use it to obtain volatility structure using option data and compare it with VIX index, 30-day historical volatility measure and close of day implied volatility of the options. The dataset for volatility structure is obtained by taking the call-options with maximum traded volume on a given quote date. Thereafter the adaptive state estimation strategy is used similar to option pricing forecasting framework.  \par
Though this procedure is computationally expensive, hence will be difficult to use on high-frequency data, it is feasible for online estimation in intraday data. For the prediction step of filters, the underlying price in BS model is modelled as geometric Brownian motion.

\section{Experimental Results and Analysis}
We test the accuracy of the 3 filters: EKF, UKF, PF on the training dataset. We compare the result's of the 3 filters to the average and best-case switching strategy based adaptive state estimation technique proposed in our research. They will be referred to as AAF and ABF respectively hereafter. The measure of performance used is the deviation of estimated signal from the observed market option prices. The measure is represented as:
\begin{equation} 
{RMSE}=(\frac{1}{T}\sum_{t=1}^{T}\frac{(c_t - BS(\widehat{v_t}, \widehat{r_t}))^2}{K})^{\frac{1}{2}}  
\end{equation}
where $\widehat{v_t},\widehat{r_t}$ are the underlying volatility and risk states at time $t$. $c_t$ is the real option price. \par

\begin{figure*}[p]
\includegraphics[width=\textwidth, height=4 cm]{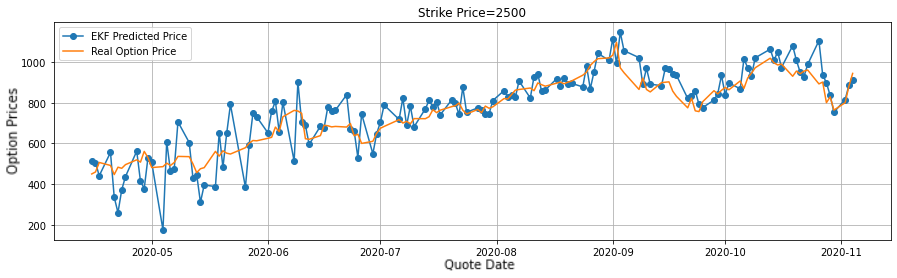}
\vspace*{\floatsep}% https://tex.stackexchange.com/q/26521/5764
\includegraphics[width=\textwidth, height=4 cm]{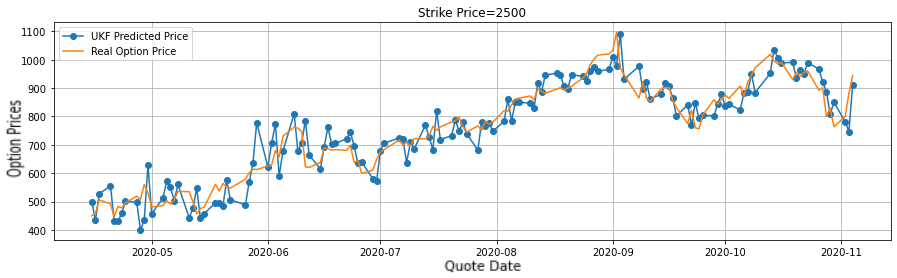}
\vspace*{\floatsep}% https://tex.stackexchange.com/q/26521/5764
\includegraphics[width=\textwidth, height=4 cm]{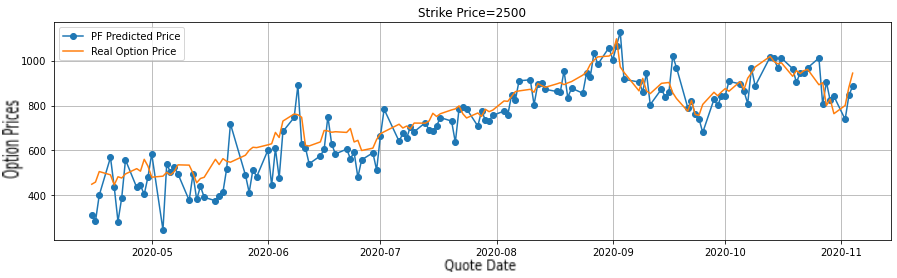}
\vspace*{\floatsep}% https://tex.stackexchange.com/q/26521/5764
\includegraphics[width=\textwidth, height=4 cm]{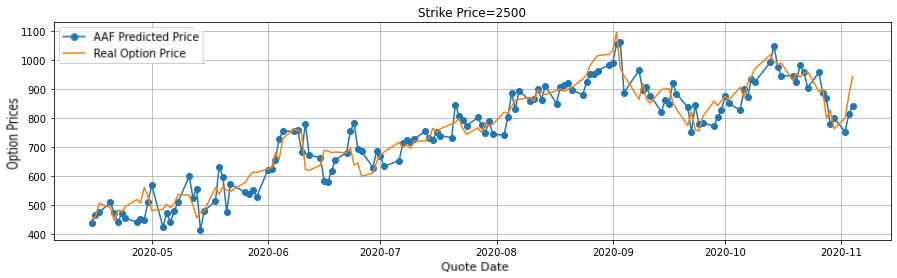}
\vspace*{\floatsep}% https://tex.stackexchange.com/q/26521/5764
\includegraphics[width=\textwidth, height=4 cm]{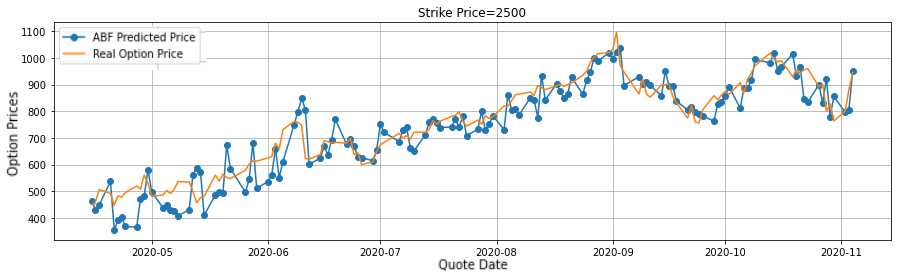}
\caption{Price Prediction for various filters for several Quote Dates}
\centering
\end{figure*}

\begin{figure*}[htp]
\includegraphics[width=\textwidth, height=5cm]{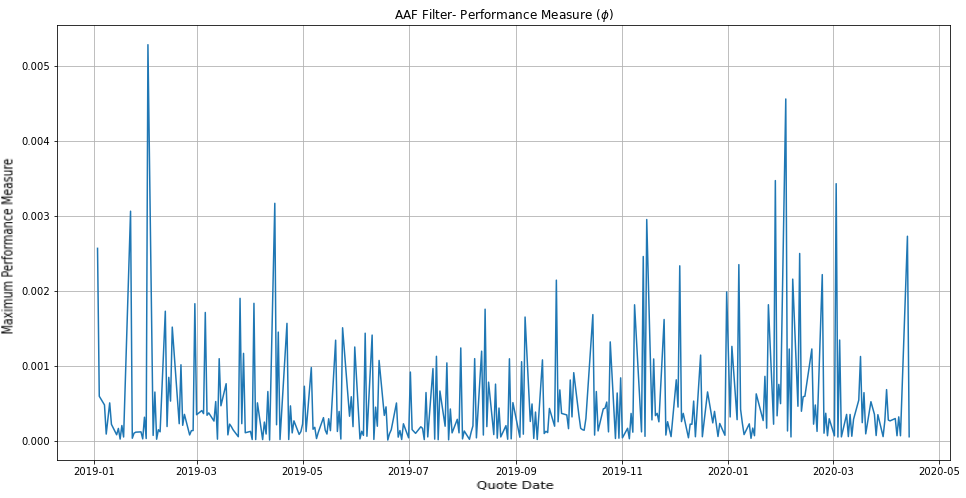}
\caption{Maximum Performance Measure computed for AAF}
\end{figure*}

\begin{figure*}[p]
\includegraphics[width=\textwidth, height=4 cm]{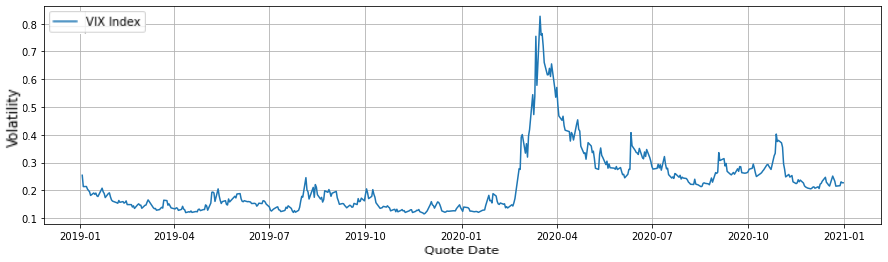}
\vspace*{\floatsep}% https://tex.stackexchange.com/q/26521/5764
\includegraphics[width=\textwidth, height=4 cm]{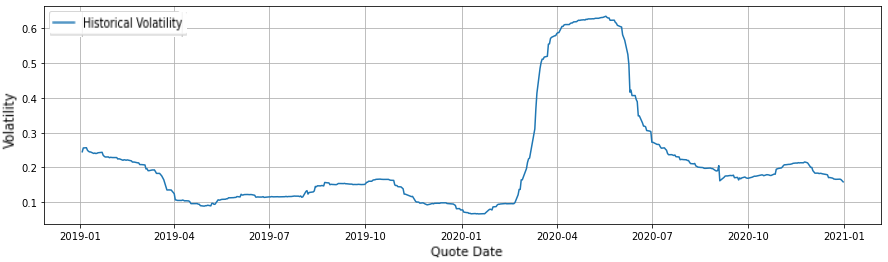}
\vspace*{\floatsep}% https://tex.stackexchange.com/q/26521/5764
\includegraphics[width=\textwidth, height=4 cm]{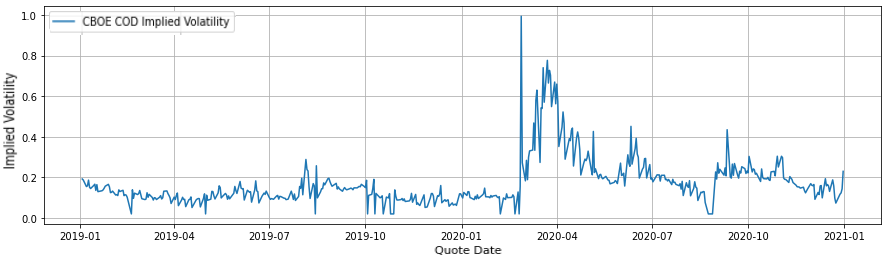}
\vspace*{\floatsep}% https://tex.stackexchange.com/q/26521/5764
\includegraphics[width=\textwidth, height=4 cm]{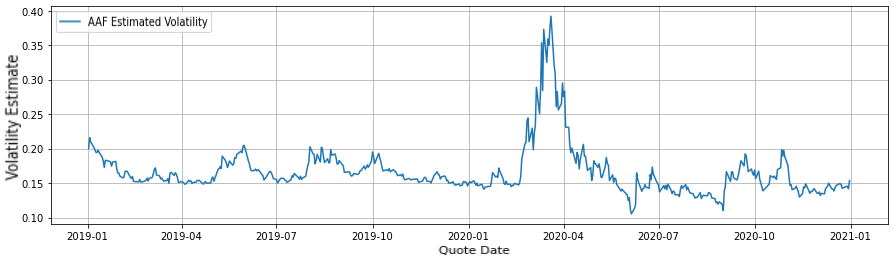}
\caption{Volatility Cmparison of AAF estimated volatility with VIX and Historical Volatility}
\centering
\end{figure*}

\begin{table}
\begin{center}
\begin{tabular}[t]{ |p{1.7cm}||p{1.5cm}|p{1.5cm}|p{1.5cm}| }
 \hline
 \multicolumn{4}{|c|}{Forecasting RMSE}\\
 \hline
 & K=2000 & K=2500 & K=3000\\
 \hline
 EKF   & 0.040    & 0.036 &  0.020\\
 \hline
 UKF &   0.033  & 0.027   & 0.019\\
 \hline
 PF &   0.031  & 0.031   &0.017\\
 \hline
 AAF &   0.028  &0.020 & 0.019\\
 \hline
 ABF &   0.030  & 0.024   & 0.018\\
 \hline
\end{tabular}
\end{center}
\caption{RMSE for Price Forecasting on Test Set}
\label{table:ta}
\end{table}

\begin{table}
\begin{center}
\begin{tabular}[t]{ |p{1.9cm}||p{1.5cm}|p{1.5cm}|p{1.5cm}| }
 \hline
 \multicolumn{4}{|c|}{Filter Frequency}\\
 \hline
 & EKF & UKF & PF\\
 \hline
 AAF   & 56    & 164 &  242 \\
 \hline
 ABF Volatility &   94  & 148   & 220\\
 \hline
 ABF Risk &   121  & 72   & 269\\
 \hline
\end{tabular}
\end{center}
\caption{Frequency of filters used for individual methods}
\label{table:ta}
\end{table}

As mentioned previously, one time step ahead forecasts are considered as they prevent the problem of cumulative errors from the previous period for out-of-sample forecasting. Since the proposed strategy considers the most optimal filter at each time step based on performance measure derived from PCRLB, it gives much better results than individual filters in terms of RMSE measure.

Table (1) displays the forecasting performance of the individual EKF model, UKF model and PF model and the proposed AAF and ABF filter. Performance was evaluated on the 3 call options. Figures (1) plots the predicted results of the various filters on option of K=2500. 
The results in Table (1) and Figure (1) indicate that the proposed adaptive state estimation technique perform better than the individual filter on option price forecasting. The technique takes the advantage of each filter and selects it adaptively according to price and underlying state movement. Fig (2) shows the maximum Performance metric $(\phi)$ computed for AAF. Table (2) gives the frequency of filter used for each strategy. Analysis shows that PF was the most used filter followed by UKF and lastly EKF for overall state and individual volatility movement. For risk, EKF was selected more than UKF, providing an insight that EKF can be suitable when a risk-free movement for an option contract is studied. \par

The technique also replicates the volatility structure of $S\&P 500$ well. As seen in Fig (3) having similar structure to VIX, historical volatility and implied volatility. The volatility estimated is mostly lower than other volatility sources.  \par
These results indicated that adaptive state estimation technique capture the nonlinear pattern of option prices in a much better way than individual filters. The technique has the flexibility to use as many Bayesian filters with the advantage of parallel computation.

\section{CONCLUSIONS}

In this paper, we use an adaptive state estimation technique for option price forecasting which takes into account the benefits of a combination of sub-optimal Bayesian estimators under appropriate non-linear model dynamics. The most optimal state estimated individually for the various filters is selected at each time step based on maximum PCRLB based performance measure, computed through a particle filter based approach. The volatility and risk were taken as the state and the option prices were modelled through Black-Scholes formula with GARCH(1,1) formulation for volatility structure.  \par
We compared the performances of the proposed technique with the individual filters in the set for estimating underlying option prices forecasting performance. It performed much better as compared to the individual filters in error metrics like RMSE. Though both measures of state covariance matrix and PCRLB are approximations, the work done provides a flexible and efficient modelling framework for estimating option prices hidden state and forecasting future prices.

\section{Future Work}
The framework described above have a large scope of modifications and extensions to other field's of finance and otherwise. With respect to the problem of volatility modelling and estimation of the underlying state of option prices for price forecasting, more advanced stochastic volatility models can be considered. Additional non-linear filtering approaches can also be added for switching strategy for adaptive state estimation. Using technique used in [7], [8], the filtering results can also be added with non-linear residuals prediction by regression techniques like SVM, Neural Networks or Boosting techniques. Lastly, our strategy has a major drawback of high time complexity, preventing its use on high-frequency trading and hedging methods. Several mathematical optimizations has to be made to reduce this drawback.

\section{Acknowledgement}
This work was supported by Professor Swanand Khare, Department of Mathematics, Indian Institute of Technology, Kharagpur. I would like to thank him for his helpful insights on concepts of Bayesian Inference, which was essential in the progress of the project.

%%%%%%%%%%%%%%%%%%%%%%%%%%%%%%%%%%%%%%%%%%%%%%%%%%%%%%%%%%%%%%%%%%%%%%%%%%%%%%%%

%%%%%%%%%%%%%%%%%%%%%%%%%%%%%%%%%%%%%%%%%%%%%%%%%%%%%%%%%%%%%%%%%%%%%%%%%%%%%%%%

\end{document}